\documentclass[%
 reprint,
 amsmath,amssymb,
 aps,
 pre,
 floatfix,
]{revtex4-1}

\usepackage{graphicx}
\graphicspath{{images/}}
\usepackage{dcolumn}
\usepackage{bm}

\newcommand{\comb}[2]{C_{#1}^{#2}}
\newcommand{\la}{\langle}
\newcommand{\ra}{\rangle}
\DeclareMathOperator{\var}{var}

\begin{document}

\title{Acceptance rate is a thermodynamic function in local Monte Carlo algorithms}

\author{Evgeni Burovski$^{1}$}
\author{Wolfhard Janke$^{2}$}
\author{Maria Guskova$^{1}$}
\author{Lev Shchur$^{1,3}$}
 \email{lev@landau.ac.ru}
\affiliation{%
 $^1$ National Research University Higher School of Economics,  101000 Moscow, Russia \\
 $^2$ Institut  f\"ur Theoretische Physik, Universit\"at Leipzig, IPF 231101, 04081 Leipzig, Germany \\
 $^3$ Landau Institute for Theoretical Physics, 142432 Chernogolovka, Russia 
 }%

\date{\today}

\begin{abstract}
We study properties of Markov chain Monte Carlo simulations of classical spin models with local updates. We derive analytic expressions for the mean value of the acceptance rate of single-spin-flip algorithms for the one-dimensional Ising model. We find that for the Metropolis algorithm the average acceptance rate is a linear function of energy. 
We further provide numerical results for the energy dependence of the average acceptance rate for the 3- and 4-state Potts model, and the XY model in one and two spatial dimensions.
In all cases, the acceptance rate is an almost linear function of the energy in the critical region. The variance of the acceptance rate is studied as a function of the specific heat. While the specific heat develops a singularity in the vicinity of a phase transition, the variance of the acceptance rate stays finite.
\end{abstract}

\maketitle

\section{Introduction} 
\label{sec-intro}

One of the most challenging problems in high-performance computing (HPC) is using the full strength of supercomputers. It is very demanding, however, to efficiently use all the power of a supercomputer in a single run. The main barrier is that most of the currently available algorithms do not scale well on the complex multi-node, multi-core, and multi-accelerator hybrid architectures which are the dominant today and will be dominant for the nearest future. Hence the computation must be divided into millions of tasks to be scheduled on individual cores. It is thus of crucial importance to develop new, fully scalable algorithms, new programming techniques, and new methods to build programs which can efficiently use the power of supercomputers.

Markov chain Monte Carlo methods ~\cite{Berg,BinderLandau} are among the prime approaches of supercomputer simulations in physics, chemistry, and materials sciences. A Monte Carlo (MC) simulation is a sequence of local updates of microscopic degrees of freedom, and the overall performance of a simulation strongly depends on these local elementary updates~\cite{BinderLandau}. 
Two features of the family of local MC algorithms are worth noting: first,
simulations based on these algorithms are naturally embarrassingly parallel,
which makes them even more attractive for massively parallel computations;
second, they are applicable in the presence of external fields and/or competing
ferro- and antiferromagnetic interactions, where more sophisticated schemes
(e.g., cluster updates) break down.

A program of the US DOE, Office of Science, entitled the Innovative and Novel 
Computational Impact on Theory and Experiment (INCITE)~\cite{INCITE} awarded on 
average 60 projects per year, from which 30 percent are in physics, 28 percent in 
engineering, 15 percent in materials sciences, 9 percent in earth sciences, and 
7 per cent in chemistry, etc. The average number of projects which used MC 
methods is 5 per year in the last ten years. 

In the core of MC methods is the Metropolis algorithm~\cite{Metropolis} known for more than 60 years.  What we find surprising is that a careful  analysis of an ``old'' method could bring completely new knowledge on the computations. Namely, we note that the acceptance rate of an MC simulation is a thermodynamic function which displays a well-defined temperature dependence.
These findings provide a deeper understanding of the algorithm, and we believe that our observation may influence further improvements of MC algorithms in general, and their parallel scalable versions.

A Markov process is defined by its set of transition probabilities between microscopic states, which must obey a balance condition \cite{Berg,BinderLandau}.
Efficiency of an MC process is governed by several factors: the typical acceptance rate (i.e., the fraction of states in the Markov chain which differ from the previous state), the computational complexity of an elementary update, and the autocorrelation time of the process. Different choices for both trial moves and their transition probabilities are possible~\cite{BinderLandau, Janke2008}.
A rigorous theorem claims that a specific choice of elementary updates --- which is in fact a global update, attempting to update all degrees of freedom at once by drawing random increments from a Gaussian distribution --- maximizes the efficiency
if the mean value of the acceptance rate is tuned to the special value 0.234~\cite{Baesian, Baesian2}. However, such processes have unfavorable correlation properties~\cite{PotterSwendsen2013}, and are not suitable for MC simulations of physical systems with a large number of degrees of freedom.

We now make the following observation: the mean value of the acceptance rate of an MC simulation using a given set of transition probabilities has a well-defined temperature dependence. Therefore, it can be viewed as a thermodynamic function of the model under the MC dynamics, on par with other thermodynamic functions, e.g., the energy. A natural question is then: what is the relation between the mean acceptance rate and other thermodynamic quantities? We find that for the one-dimensional (1D) Ising model, the acceptance rate of the Metropolis algorithm \cite{Metropolis} is a \emph{linear} function of energy. An immediate question is then whether this linear relation is a one-off artifact of the Metropolis MC dynamics for the 1D Ising model, or whether it generalizes for other related models and MC algorithms.

The rest of the paper is organized as follows. In Sec.~\ref{sec-models} we briefly describe the models for which we calculate analytically or compute numerically the acceptance rate. In Sec.~\ref{sec-analytics} we present analytical results on the acceptance rate of the Metropolis and heat-bath algorithms for the 1D Ising model. Section~\ref{sec-simulations} contains computational results for the acceptance rate and its variance for the one-dimensional models described in Sec.~\ref{sec-models}. Next Sec.~\ref{sec-2d} presents computational results for the acceptance rate and its variance for the two-dimensional models. In Sec.~\ref{sec-discussion} we summarize our findings. Two Appendixes contain more details on the analytical calculation of the acceptance rate when applying the Metropolis and heat-bath algorithms to the 1D Ising model.

\section{Models and Update Algorithms}
\label{sec-models}

We consider several well-known classical lattice models. The  Ising model is defined by the Hamiltonian function
\begin{equation}
H = -J \sum_{\langle ij \rangle} S_i S_{j}\;,
\label{ising_S}
\end{equation}
where the coupling constant $J > 0$,
$\langle ij \rangle$ denotes nearest-neighbor pairs,
and $S_i = \pm 1$ are Ising spins, located 
at the sites of a $d$-dimensional lattice of linear size $L$ 
(and volume $V = L^d$) with periodic boundary conditions.

In the  $q$-state Potts model, spins can take $q$ possible values, 
$S_i\in {1, \ldots,q}$ \cite{Wu}:
\begin{equation}
H = -J\sum_{\langle ij \rangle} \delta(S_i, S_{j})\;,
\label{potts_S}
\end{equation}
where the coupling constant $J>0$ and $\delta(S_i, S_{j})$ is the Kronecker delta symbol, which equals one
whenever $S_i=S_{j}$, and zero otherwise.

Finally, we consider the XY model, defined by \cite{Kosterlitz}:  
\begin{equation}
H = -J\sum_{\langle ij \rangle} \cos{(S_i-S_{j})}\;,
\label{XY_S}
\end{equation}
where the coupling constant $J>0$ and $S_i$ are continuous variables,
$S_i \in [0, 2\pi)$.

MC simulations provide a way of
studying models \eqref{ising_S}-\eqref{XY_S} in thermodynamic equilibrium.
An MC simulation constructs an ergodic random walk in the configuration
space of a model,
$$
\cdots\to \mu \to \nu \to \cdots \;,
$$
which generates the equilibrium Gibbs distribution of a model as its stationary
distribution \cite{BinderLandau}. For local updating schemes, successive
configurations $\mu$ and $\nu$ only differ by the value of a single spin. 

An elementary update of the local Metropolis algorithm \cite{Metropolis} 
for the Ising model proceeds in two steps: (i) select a random site
and (ii) flip its spin, $S_i \to -S_i$, with the probability 
\begin{equation}
p(\mu\to\nu) = \min(1, e^{-\beta\Delta E}) \;,
\label{metropolis_p}
\end{equation}
where $\beta$ is the inverse temperature and $\Delta E = E_\nu - E_\mu$ the energy 
difference between the updated and original states \cite{Janke2008}. The
generalization to models with more than two states per spin is straightforward:
(ii) is simply replaced by $S_i \to \widetilde{S}_i$, where $\widetilde{S}_i$ is 
any admissible spin value.

The heat-bath algorithm for the Ising model differs from the Metropolis algorithm 
only in that a spin-flip update is accepted with the probability \cite{Janke2008}
\begin{equation}
p(\mu\to\nu) = \frac{e^{-\beta E_\nu}}{e^{-\beta E_\nu} + e^{-\beta E_\mu}}\;.
\label{heat_bath_p}
\end{equation}
This can be recast into the form
\begin{eqnarray}
p(\mu\to\nu) &=& \frac{e^{-\beta \Delta E/2}}{e^{-\beta \Delta E/2} + e^{\beta \Delta E/2}}\nonumber\\
&=& \frac{1}{2}\left[1-\tanh(\beta \Delta E/2)\right]\;,
\label{heat_bath_p2}
\end{eqnarray}
which is the general Glauber update rule \cite{Glauber} that can also be applied
to models with more than two states per spin and a generalized update proposal
$S_i \to \widetilde{S}_i$.
Note that only for the Ising model with two states per spin, the heat-bath process 
coincides with the Glauber dynamics \cite{Glauber}. In the general case, the heat-bath 
process in a strict sense (there is some confusion with the notation in the literature) 
involves {\em all\/} possible spin values and is hence more complicated.

\section{Acceptance rates of MC simulations of the 1D Ising model}
\label{sec-analytics}
To calculate the expected value of the acceptance probability of an MC simulation
of the 1D Ising model we first convert Eq.\ \eqref{ising_S} to bond
variables \cite{Suzuki1972, Mueller2017}.

\subsection{Bond representation}
\label{subsec:ising_bonds}

In one dimension Eq.\ \eqref{ising_S} takes the form
\begin{equation}
H = -J \sum_{i=1}^{L} S_i S_{i+1}\;,
\label{ising_S_1D}
\end{equation}
where the indices in \eqref{ising_S_1D} are taken modulo $L$,
i.e., the term $S_L S_{L+1}$ is understood as $S_L S_{1}$. 

To calculate the expected values of the acceptance probability of a MC simulation
of the 1D Ising model \eqref{ising_S_1D} for the Metropolis and heat-bath updates,
we first convert the model \eqref{ising_S_1D} to a bond representation.
We define for a bond connecting sites $i$ and $i+1$ the
'charge' \cite{Suzuki1972, Mueller2017},
\begin{equation}
Q_i = \frac{1}{2}\left(S_i S_{i+1} + 1 \right)\;,
\label{q_def}
\end{equation}
which takes values of 0 (for $S_i \neq S_{i+1}$) and 1 (for $S_i = S_{i+1}$ ).
In this representation, Eq.\ \eqref{ising_S_1D} takes the form
\begin{equation}
H = - 2 J \sum_{i=1}^L Q_i + J L\;,
\label{ising_Q}
\end{equation}
where the sum is taken over the \emph{bonds} of the lattice. With periodic boundary
conditions, the number of bonds equals the number of sites of the lattice.
This way, the state space of the model \eqref{ising_S_1D} is spanned by a
collection of $L$ integers $Q_i = \left\{0, 1\right\}$, subject to the
constraint: the parity of $\sum_{i=1}^L Q_i$ is the same as the parity of the
number of bonds. We take $L$ to be even throughout, so that 
$$\sum_{i=1}^L Q_i \text{\qquad is even.} 
$$

The partition function corresponding to Eq.\ \eqref{ising_Q} then reads
\begin{equation}
Z = 2 x^{-L/2} \sum_{l = 0}^{L/2} \comb{L}{2l} x^{2l}\;,
\label{Z}
\end{equation}
where $x \equiv e^{2\beta J}$ and $\beta$ is the inverse temperature. 
The summation runs over the values of $\sum_i Q_i = 2l$, and the binomial
coefficient, $\comb{L}{2l}$, counts the number of ways of distributing $2l$
values of $Q=1$ over $L$ bonds.
The factor of 2 accounts for a double-counting of the representation \eqref{q_def}:
each value of $Q_i$ can be realized by two possible combinations of $S_i$ and $S_{i+1}$
(e.g., $Q_i = 0$ means that either $S_i=-1$ and $S_{i+1} = 1$ or vice versa).

Performing the summation in \eqref{Z} we obtain
\begin{equation}
Z = x^{-L/2} \Bigl[ (x+1)^L + (x-1)^L \Bigr]\;,
\label{Z_2}
\end{equation}
which agrees with the standard result \cite{Baxter}.

\subsection{Acceptance rate of the Metropolis algorithm}

We use the bond representation \eqref{ising_Q} to calculate the acceptance rates for the Metropolis and the heat-bath algorithms.
In this section, we only state the main results and relegate the details of calculations to the Appendix.

We start with the Metropolis update \eqref{metropolis_p}. Denoting the
expected value of the acceptance probability by $R$, the expected value of the \emph{rejection} probability is
\begin{equation}
1 - R = \frac{x^2-1}{Z} %
\Bigl[ (x+1)^{L-2} + (x-1)^{L-2}\Bigr] x^{-L/2} \;.
\label{expect_2}
\end{equation}
In the thermodynamic limit, $L\rightarrow \infty$, the second term in brackets is negligible, and Eq.\ \eqref{expect_2} simplifies to
\begin{equation}
R = \frac{2}{x+1}\;.
\label{acpt_prob}
\end{equation}
We now compare Eq.\ \eqref{acpt_prob} to the thermodynamic mean value of the internal energy of the system, $E$. Using the partition function \eqref{Z_2}, we obtain in the thermodynamic limit the standard result \cite{Baxter}
\begin{equation}
\varepsilon = -\frac{x-1}{x+1}\;,
\label{energy}
\end{equation}
where the reduced energy density $\varepsilon = E / J L$.

Comparing Eqs.\ \eqref{acpt_prob} and \eqref{energy}, we find
\begin{equation}
R = 1 + \varepsilon \;,
\label{R_vs_e}
\end{equation}
i.e., the expected value of the acceptance probability is a {\it linear function} of the energy. In fact, relation \eqref{R_vs_e} holds for all values of $L$, see Appendix A. 

\subsection{Acceptance rate of the heat-bath algorithm}

For the expected value of the  acceptance probability 
$R$ 
of the heat-bath update \eqref{heat_bath_p} we find in the thermodynamic limit $L\rightarrow \infty$
\begin{equation}
R = \frac{x}{1 + x^2}\;.
\label{acpt_HB-of-x}
\end{equation}
Comparing to \eqref{energy}, we have
\begin{equation}
R = \frac{1}{2} \frac{1 - \varepsilon^2}{1 + \varepsilon^2} \;,
\label{acpt_HB}
\end{equation}
which approaches the linear behavior $(1+\varepsilon)/2$ for $1 + \varepsilon \ll 1$
(cf.\ Fig.\ \ref{I1M_HB_ER}).
Details can be found in Appendix B.

\section{Simulation results in 1D}
\label{sec-simulations}

In this section we first verify the analytical results for the
Metropolis and heat-bath acceptance rates of the Ising model in
one dimension and then test the observed qualitative features 
for the other models defined in Sec.\ \ref{sec-models}. Results
for the generalization to two dimensions will be presented in the
next section.

\subsection{First moments of the energy and acceptance rate}

We performed MC simulations of the 1D Ising model \eqref{ising_S_1D} using Metropolis
updates \eqref{metropolis_p} and heat-bath updates \eqref{heat_bath_p} for
temperatures ranging from $T/J = 0.2$ to $10$.
We used $N_T=10^6$ MC steps (MCS) for thermalization
and collected statistics over $N_A=10^7$ MCS. Here an MCS is defined
as $L$ elementary update attempts for a chain with $L$ spins. We first 
focus on the collected
statistics for the total energy of the system, $E$, and
the acceptance rate, $R$, which we specifically define as the ratio of the
numbers of accepted and attempted elementary updates.

\begin{figure}[tb]
\includegraphics[width=.99\columnwidth]{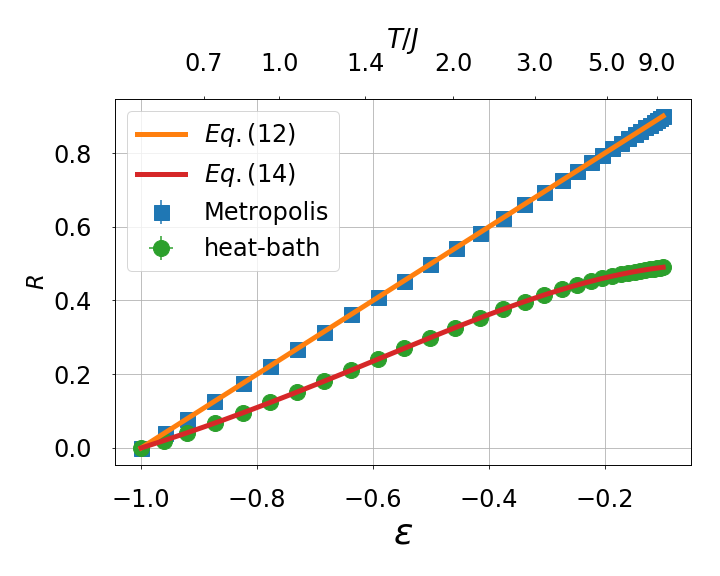} 
\caption{Average acceptance rates of the Metropolis updates \eqref{metropolis_p} and the heat-bath updates \eqref{heat_bath_p} for the 1D Ising model \eqref{ising_S_1D}. Simulation were done for $L=512$ spins with periodic boundary conditions. Symbols are simulation results, with error bars shown at all points. Solid lines are predicted relations \eqref{R_vs_e} and \eqref{acpt_HB}. See text for discussion.}
\label{I1M_HB_ER}
\end{figure}

Figure \ref{I1M_HB_ER} shows the relation between the mean values of the
acceptance rates of the MC process and the reduced energy density, $\varepsilon$,
for a chain of $L = 512$ spins and periodic boundary conditions.
The results of the MC simulations agree with  Eqs.\ \eqref{R_vs_e} and \eqref{acpt_HB} 
in the whole range of energies (hence, temperatures). 
We note that for the heat-bath algorithm the dependence of 
$R$ 
on the reduced energy density is approximately linear in a wide
range of temperatures: the relative difference between 
$R$ and its
linear approximation is below 10\% for $T/J < 1.1$.

It is instructive to compare the behavior of local MC algorithms for related classical 
spin models. We performed MC simulations of the 3- and 4-state Potts models \eqref{potts_S},
and the classical XY model \eqref{XY_S} in one dimension using the Metropolis and 
Glauber algorithms.
There are several ways of organizing the local updates. We take the simplest possible prescription: we select a spin $S_i$ at random, and then draw a proposed value $\widetilde{S}_i$ for the update $S_i \to \widetilde{S}_i$ from a uniform discrete distribution of $q$ values for the Potts model, and from a uniform distribution on $[0, 2\pi)$ for the XY model.
In these simulations we use $N_T=10^4$ MCS for thermalization and $N_A=10^5$ MCS for the averaging.

\begin{figure*}[bt]
\begin{tabular}{cc}
\includegraphics[width=.99\columnwidth]{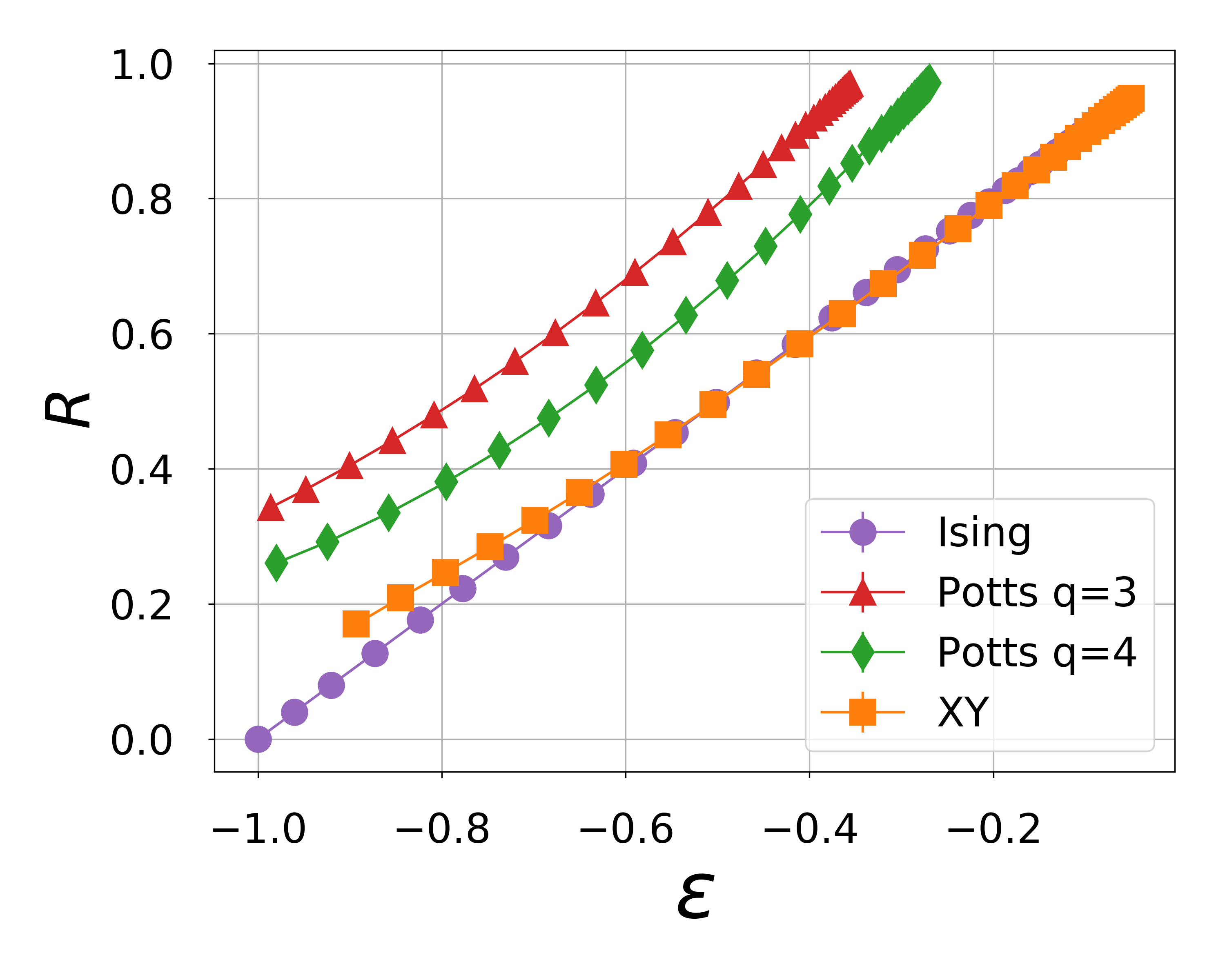} &
\includegraphics[width=.99\columnwidth]{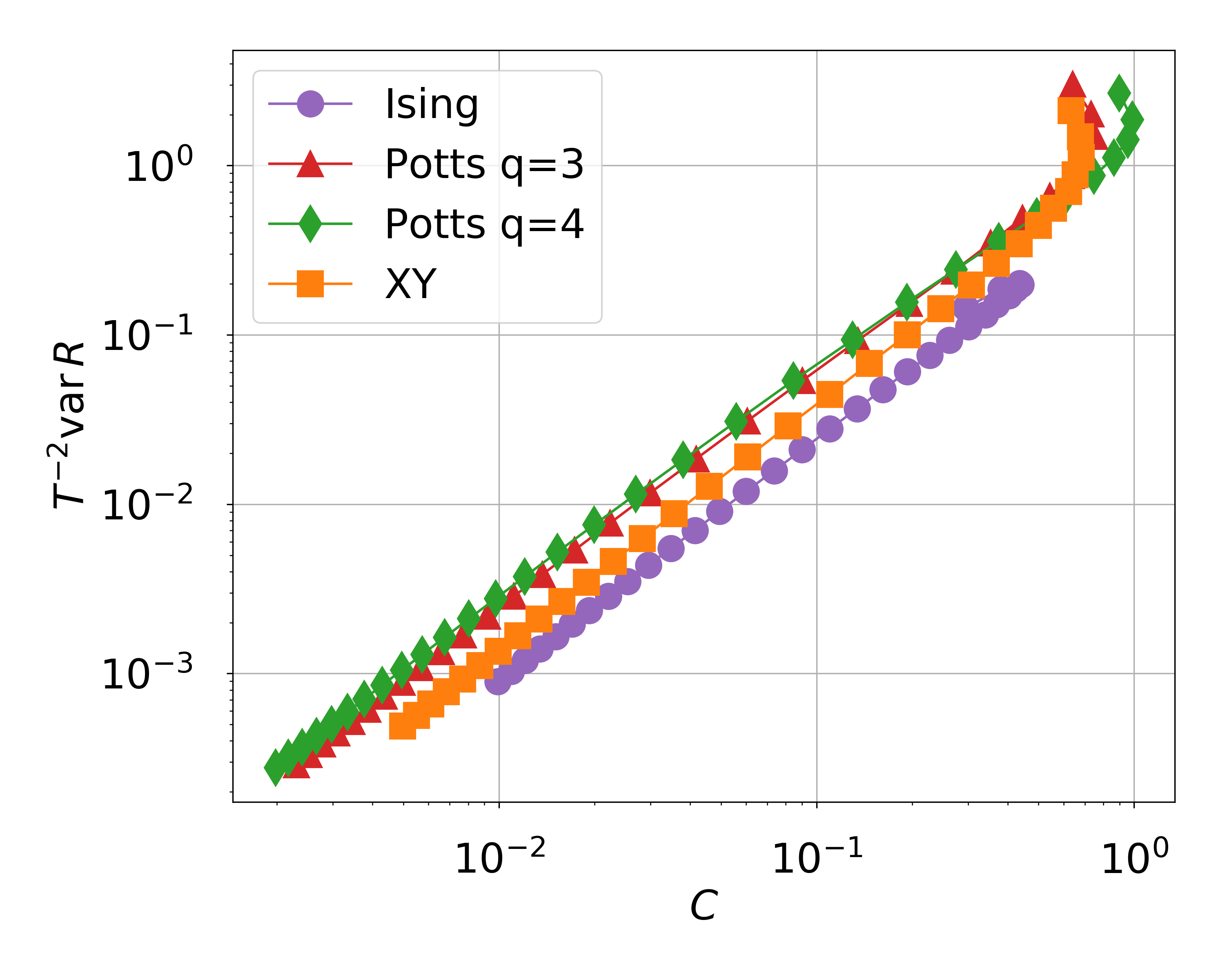} \\
\end{tabular}
\caption{(a) Average acceptance rates of the Metropolis updates \eqref{metropolis_p} versus energy for 1D models. (b) Variance of the acceptance rate versus heat capacity on a log-log scale. Simulations were done for 
the Ising model (circles), the Potts model with $q=3$ (triangles) and $q=4$ (diamonds), and the XY model (squares),
using chains of $L=512$ spins with periodic boundary conditions.
Symbols are simulation results with error bars, lines are to guide the eye. 
}
\label{ALL1M_HB_ER}
\end{figure*}

\begin{figure*}[bt]
\begin{tabular}{cc}
\includegraphics[width=.99\columnwidth]{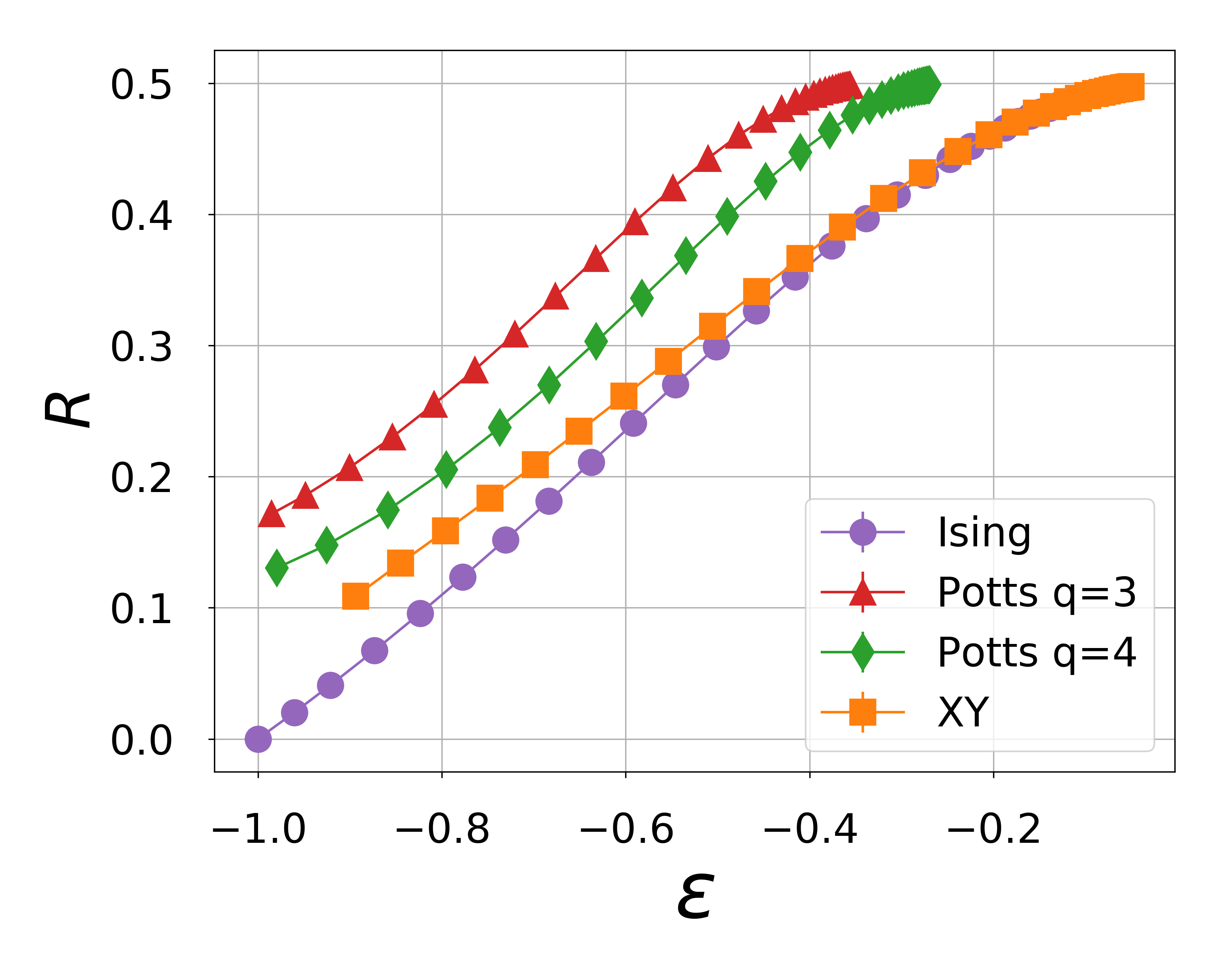} &
\includegraphics[width=.99\columnwidth]{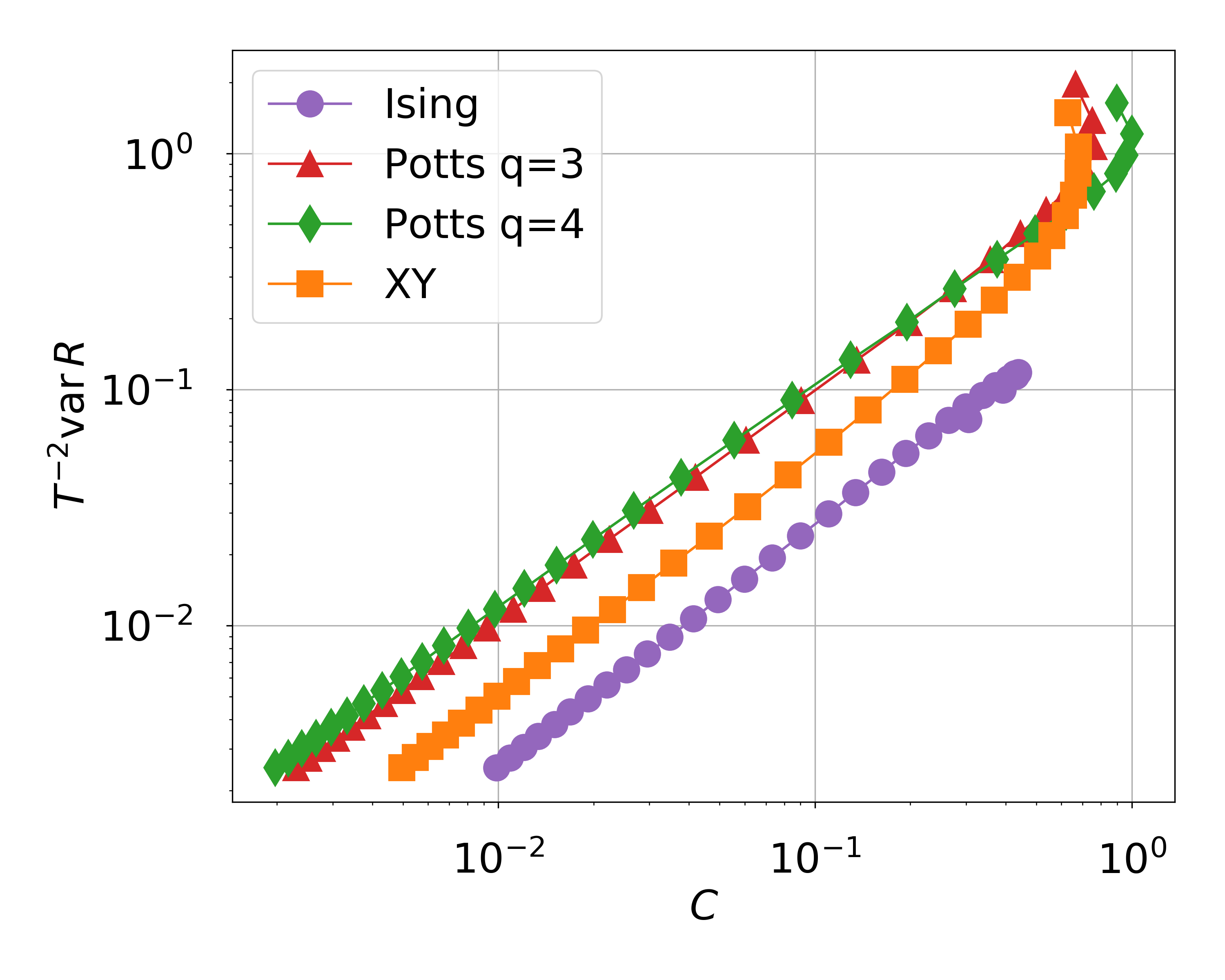} \\
\end{tabular}
\caption{(a) Average acceptance rates of the heat-bath respectively Glauber 
updates \eqref{heat_bath_p}, \eqref{heat_bath_p2} versus energy for 1D models. (b) Variance of the acceptance rate versus heat capacity on a log-log scale. Simulations were done for the Ising model (circles), the Potts model with $q=3$ (triangles) and $q=4$ (diamonds), and the XY model (squares), 
using chains of $L=512$ spins with periodic boundary conditions.
 Symbols are simulation results with error bars, lines are to guide the eye. 
}
\label{Supp_ALL1M_HB_ER}
\end{figure*}

The energy dependence of the acceptance rate for the Metropolis algorithm is summarized in Fig.\ \ref{ALL1M_HB_ER}(a). In general, the dependence turns out to be a non-linear featureless curve. The maximum difference is observed to be between the 3-state Potts and Ising models. The 4-state Potts model is closer to the Ising result, and for the XY model, the acceptance rate approaches that for the Ising model at very large temperatures, $T \gg J$.
Results for the 
Glauber updates turn out to be qualitatively similar, and we present them for one spatial 
dimension in Fig.\ \ref{Supp_ALL1M_HB_ER}(a).

\subsection{Second moments of energy and acceptance rate} %

Mean values of energy and acceptance rate are computed as first moments of samples
generated by the MC process. Given a one-to-one, monotonic relation between the
mean values, it is instructive to compare the second moments.
In general, the second moment of the reduced energy density $\varepsilon = E/JV = \la H \ra/JV$ is related to the specific-heat capacity, 
\begin{equation}
C = J\frac{d \varepsilon}{dT} = \frac{\la H^2 \ra - \la H \ra^2}{VT^2}\;,
\label{hcap}
\end{equation}
where $\la \cdots\ra$ stands for the average over the states generated by the MC process.

The variance of the acceptance rate is readily computed as the variance of a
Bernoulli process of binary decisions ($1$ if an elementary update is accepted,
and $0$ otherwise). For a Bernoulli process, the variance, $\var R$, is related
to the mean value, $R$, via $\var R = R (1 - R)$.
In the case of the 1D Ising model, given the exact results for $R$ derived 
above, hence also $\var R$ is known exactly.

Figure \ref{ALL1M_HB_ER}(b) displays the relation between the heat capacity and the
variance of the acceptance rate for the Metropolis algorithm. We scale the variance of $R$ by $T^2$ in accordance
with Eq.\ \eqref{hcap}. At lowest temperatures, $T \ll J$, the heat capacity as a
function of temperature has a maximum due to the well-known Schottky anomaly \cite{Kubo}. Because
of this, the curves in Fig.\ \ref{ALL1M_HB_ER} form arcs for $C\sim 1$. Outside
of this range, for $T > J$, the relation between second moments divided by $T^2$
is close to linear on the log-log scale for all one-dimensional models.
Results of simulations using the heat-bath respectively Glauber updates, shown
in Fig.\ \ref{Supp_ALL1M_HB_ER}(b), look qualitatively similar.

\section{Simulation results in 2D} 
\label{sec-2d}

It is instructive to compare the behavior of 1D models to higher dimensions, if
only to see whether our observations are specific to 1D or have broader applicability.
Specifically, 2D models with Hamiltonians \eqref{ising_S}--\eqref{XY_S} undergo a
phase transition at a certain temperature $T_c$, between a high-temperature
paramagnetic behavior and a low-temperature 
phase. For the Potts
models, critical parameters are known analytically \cite{Wu},
$T_c / J = 1 / \ln\left(1 +\sqrt{q} \right)$, and for the 
Kosterlitz-Thouless transition of the XY model, MC simulations
of Ref.~\cite{Weber} quote $T_c / J = 0.887(2)$.
Here the specific-heat capacity stays finite everywhere, with a 
smooth peak located about $20\%$ above $T_c/J$.
The behavior of the MC process with local updates varies significantly between
the paramagnetic phase ($T > T_c$), the low-temperature phase ($T < T_c$), and
the critical region ($T \approx T_c$) \cite{BinderLandau}.

\begin{figure*}[htb]
\begin{tabular}{cc}
\includegraphics[width=.99\columnwidth]{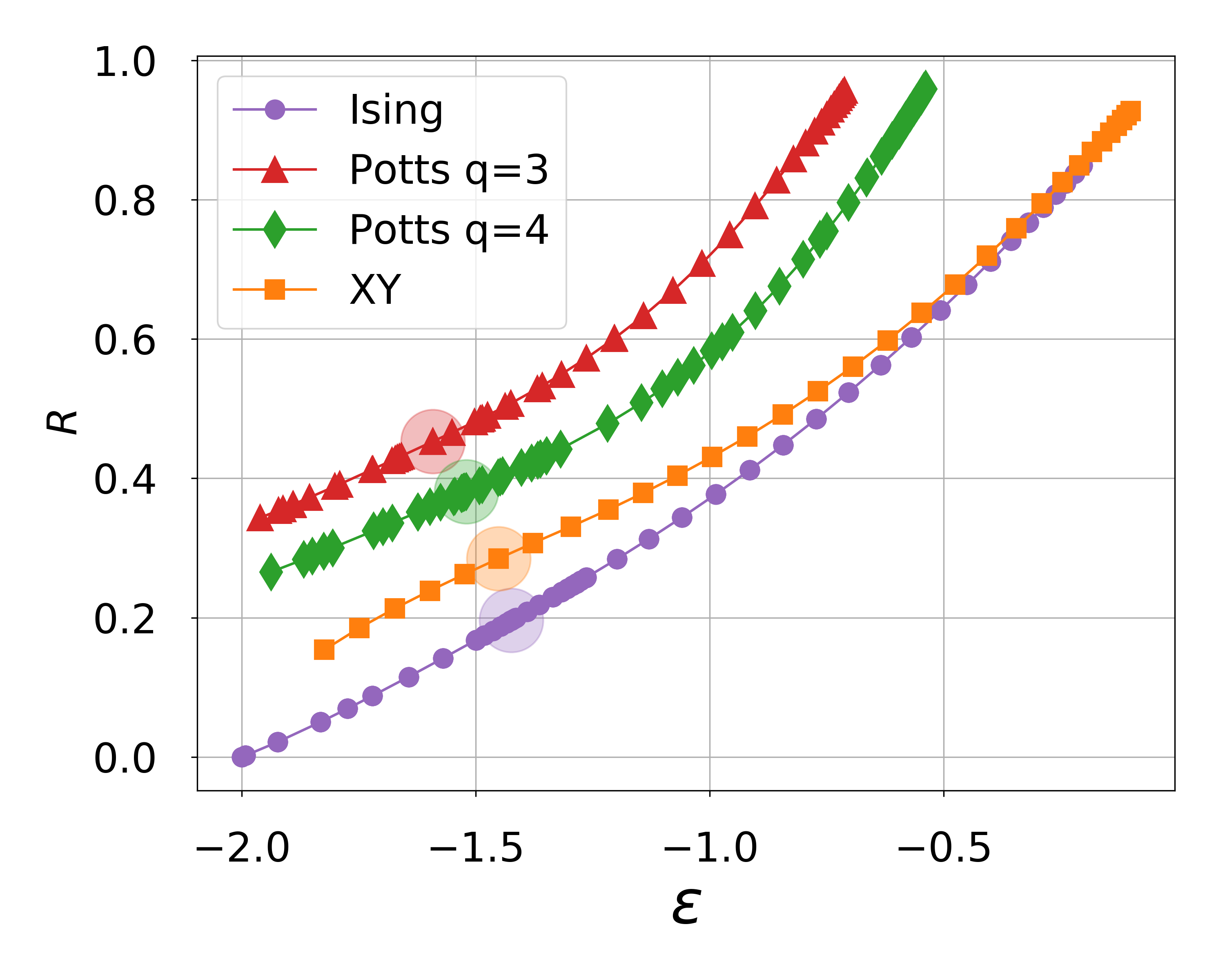} &
\includegraphics[width=.99\columnwidth]{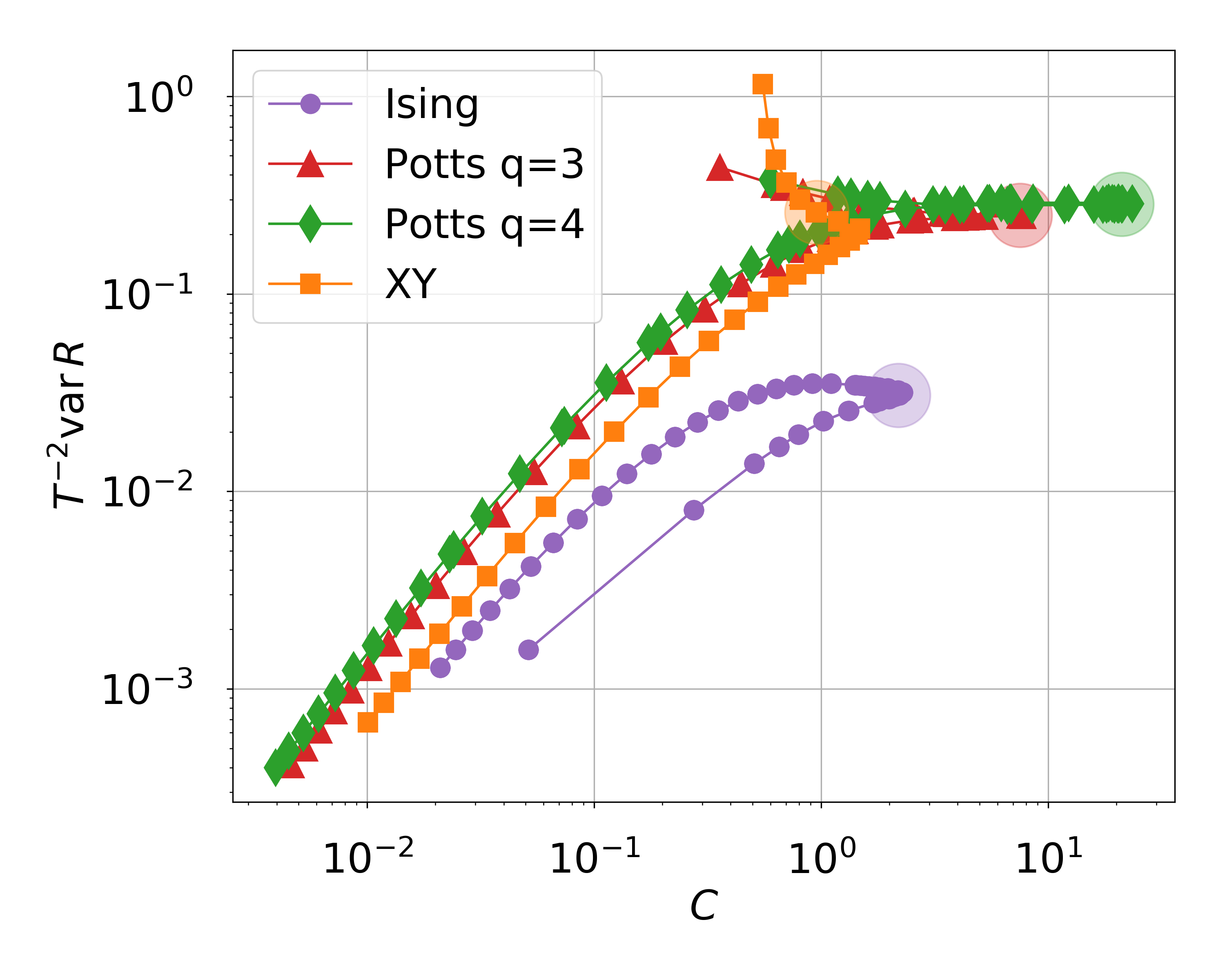} \\
\end{tabular}
\caption{(a) Average acceptance rates of the Metropolis updates \eqref{metropolis_p} for 
2D models. (b) Second moments of energy and the acceptance rate on a log-log scale. Simulations were done 
on a $64^2$ square lattice with periodic boundary conditions for the Ising model (circles), the Potts model with $q=3$ (triangles) and $q=4$ (diamonds), and the XY model (squares),
using $N_T=10^5$ MC steps (MCS) for thermalization and $N_A=10^6$ MCS for averaging.
Symbols are simulation results with error bars, lines are to guide the eye. Semi-transparent disks show the critical regions. 
}
\label{All2M_HB_ER}
\end{figure*}

Figure~\ref{All2M_HB_ER} shows results of simulations with the Metropolis algorithm
of the two-dimensional models with $64^2$ spins for temperatures between $T/J = 0.5$ to $10$.
Here we use $N_T=10^5$ MCS for thermalization and $N_A=10^6$ MCS for averaging.

The first moments of the acceptance rate and energy shown in Fig.~\ref{All2M_HB_ER}(a)
are both smooth across the phase
transition, with a relation which is close to linear in the critical region
$T \sim T_c$. 
The second moments in Fig.\ \ref{All2M_HB_ER}(b) show two clearly separate
branches, for $T > T_c$ and $T < T_c$, which join at the critical point. 
Note that while the heat capacity develops a singularity as $T\to T_c$,
the variance of the acceptance rate remains smooth and does not show any signs of divergence.
This can be explained by recalling that the heat capacity \eqref{hcap} is
by definition proportional to the variance of the nonlocal total energy,
while the acceptance rate and its variance refer to local measurements.
It is also worth noting that the relative position of the low- and high-temperature branches
for the Ising model differs from both Potts models and the XY model.

\begin{figure*}[htb]
\begin{tabular}{cc}
\includegraphics[width=.99\columnwidth]{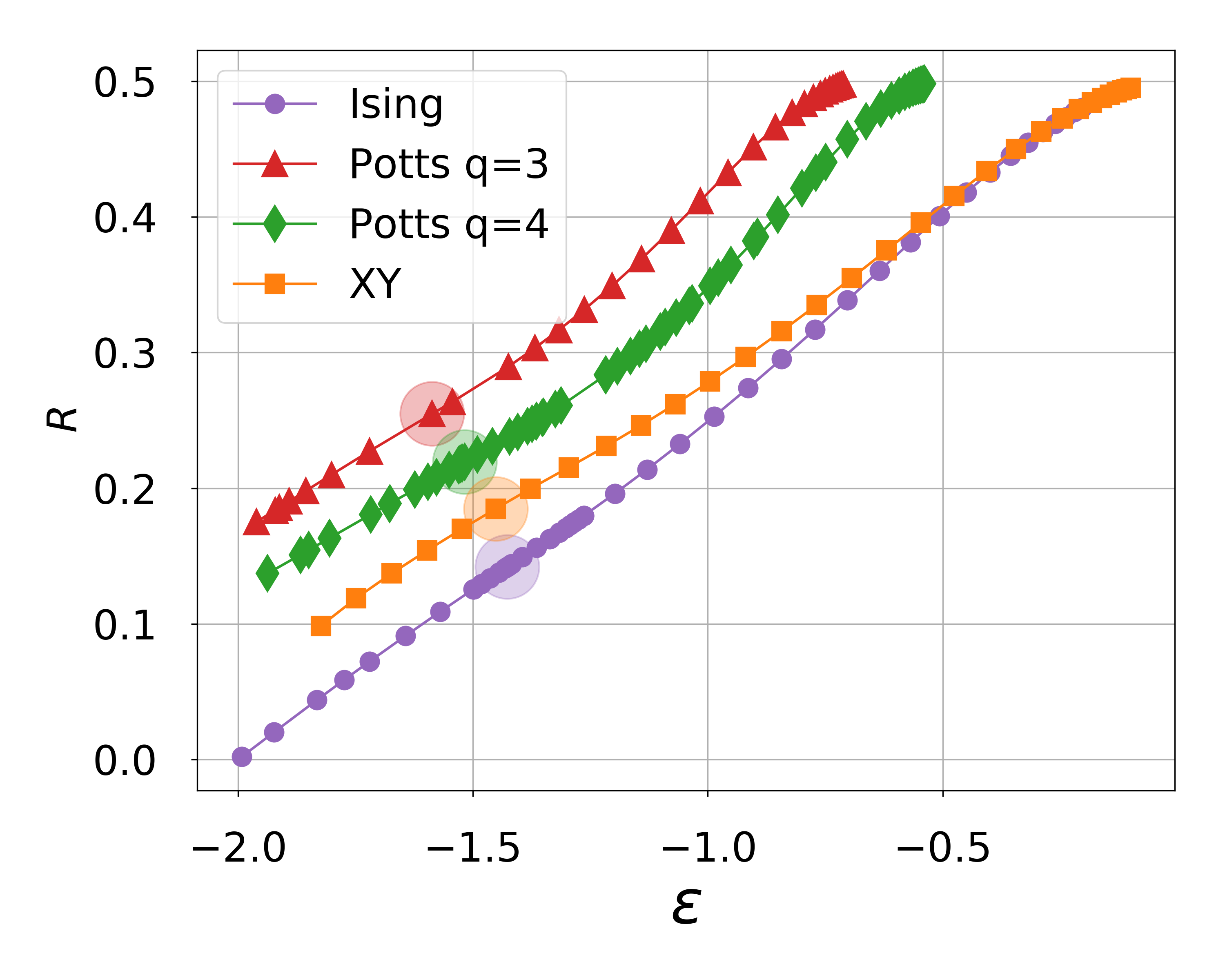} &
\includegraphics[width=.99\columnwidth]{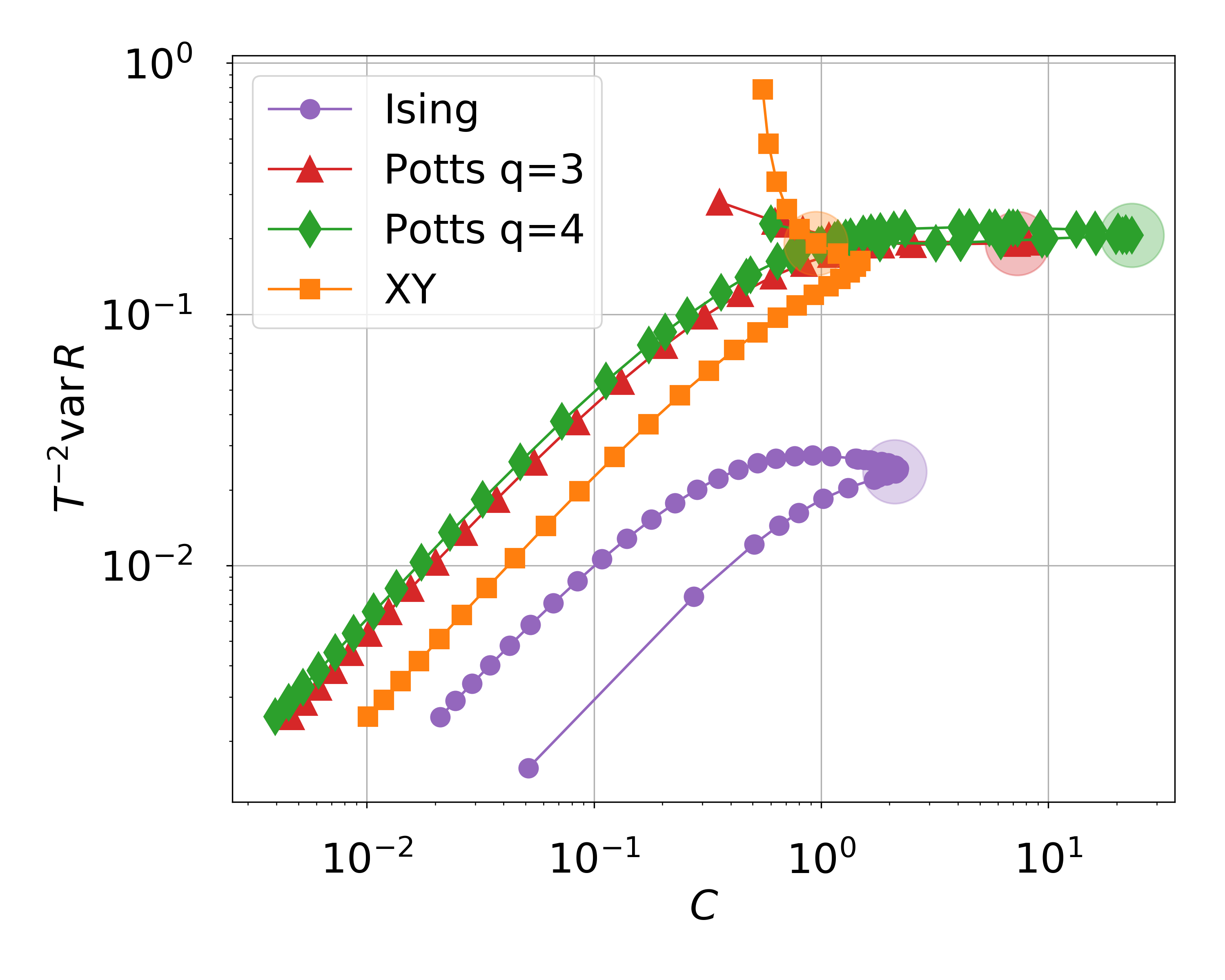} \\
\end{tabular}
\caption{(a) Average acceptance rates of the heat-bath respectively Glauber 
updates \eqref{heat_bath_p}, \eqref{heat_bath_p2} versus energy for 2D models. (b) Variance of the acceptance rate versus heat capacity on a log-log scale. Simulations were done on a square lattice with $64^2$ spins and periodic boundary conditions for the Ising model (circles), the Potts model with $q=3$ (triangles) and $q=4$ (diamonds), and the XY model (squares),
using $N_T=10^5$ MC steps (MCS) for thermalization and $N_A=10^6$ MCS for averaging.
Symbols are simulation results with error bars, lines are to guide the eye. 
Semi-transparent disks indicate the critical region of the corresponding model. 
}
\label{Supp_All2M_HB_ER}
\end{figure*}

For heat-bath respectively Glauber updates, the results are qualitatively similar to 
those using Metropolis updates, see Fig.\ \ref{Supp_All2M_HB_ER}.

We have also verified that simulations of the three- and four-dimensional models 
behave qualitatively similar to the two-dimensional models. These results will be 
detailed elsewhere.


\section{Conclusion} 
\label{sec-discussion}

Concluding, we start with the observation that the acceptance rate of an
update proposal of a Monte Carlo simulation can be regarded on par with thermodynamic
functions of a model, and thus can itself be considered a thermodynamic function
of a model under a given Monte Carlo dynamics. 

For the one-dimensional Ising model we derive analytically that the mean value of 
the acceptance rate for local Metropolis updates is a linear function of energy. 
This linear dependence turns out to be specific for this combination of the updating
scheme and the model: changing the updating algorithm to heat-bath updates changes
the functional form of the relation between the mean values of the acceptance rate 
and energy, so that the relation is only linear away from the high-temperature 
region $T \gg J$. 

We simulate several classical models in one and two spatial dimensions, 
the Ising model, the 3- and 4-state Potts models, and the XY model, and compute the
dependence of the first and second moments of the acceptance rate on the mean value
and the second moment of energy. We find that in general the relation is not linear
in a wide range of temperatures, but is close to linear in the critical region
around the transition temperature $T_c$.

Our result for the acceptance rate of the heat-bath algorithm for the one-dimensional 
Ising model can be viewed as an addition to the Glauber paper~\cite{Glauber} on the 
dynamics of the one-dimensional Ising model -- since in that case, the heat-bath 
algorithm exactly reproduces the Glauber dynamics. The acceptance rate in the Glauber dynamics \cite{Glauber} is the \emph{frequency} of the spin flips, and it is given by Eqs.\ \eqref{acpt_HB-of-x} and \eqref{acpt_HB}.

The acceptance rate can also be calculated analytically for any exactly solvable model
(e.g., one-dimensional $q$-state Potts models and the two-dimensional
Ising model), although this is not straightforward in all cases. One can compute
the acceptance rate for any local Monte Carlo algorithm. In fact, we checked that for
all models and for all dimensions for which we have code on hand,
the acceptance rate is linear in the energy close to a 
second-order phase transition, as demonstrated for some two-dimensional models
in Section V. Moreover, additional simulations show that in the vicinity of a first-order
phase transition, the acceptance rate is a linear function of energy as well.

The source codes for our Monte Carlo simulations and data analysis are available from 
Ref.~\cite{repo}.

\begin{acknowledgments}
       
E.B. and M.G. acknowledge the support of the Academic Fund Program at the National Research University Higher School of Economics (HSE), grant  No.~18-05-0024, and by the Russian Academic Excellence Project ``5-100''.
W.J. thanks the Deutsche Forschungsgemeinschaft (DFG, German Research Foundation) for support through project No.\
189\,853\,844 -- SFB/TRR 102 (project B04). L.S. acknowledges the support from the program 0033-2018-0010 of Landau Institute for Theoretical Physics and thanks Travis S. Humble (Oak Ridge National Laboratory) for the valuable discussions of the high-performance computing and INCITE program.\\ 
\end{acknowledgments}

\section*{Appendix: Acceptance rates of local Monte Carlo updates for the 
one-dimensional Ising model}
\label{sec-appendix}

In this appendix we turn our attention to the mathematical details of calculating 
the expectations of the acceptance rates of local MC updates.

\subsection{Acceptance rate of the Metropolis algorithm}
\label{subsec-metropolis_1D_ising}

Using the bond representation \eqref{ising_Q}, we note that flipping a spin
$S_i$ flips the values of two bond charges, $Q_i$ and $Q_{i-1}$. The acceptance
probabilities depend on the sum $Q \equiv Q_i + Q_{i-1}$: for $Q = 0$ or $1$,
the Metropolis update is always accepted, since $\Delta E \leqslant 0$. For $Q = 2$, the
update is accepted with probability $e^{-4\beta J} = x^{-2}$. 

Denoting the expected value of the acceptance probability by $R$, the expected
value of the \emph{rejection} probability is then
\begin{equation}
1 - R = \sum_{l=0}^{L/2} %
\left(1-x^{-2}\right) \frac{2l}{L}\frac{2l-1}{L-1} %
\frac{\comb{L}{2l} x^{2l} 2 x^{-L/2} }{Z} \;.
\label{expect_1}
\end{equation}
Here the factor $2l (2l-1) / L (L-1)$ counts the probability that, in a
configuration with $\sum_i Q_i = 2l$, for a randomly chosen site $i$ we have $Q=2$, i.e.,
$Q_i = Q_{i-1} = 1$.

The sum entering Eq.\ \eqref{expect_1} is readily computed by differentiating
twice the binomial formula 
$$
(x + 1)^L = \sum_{k=0}^L C_L^{k} x^k \;.
$$
The result is
$$
1 - R = \frac{x^2-1}{Z} %
\Bigl[ (x+1)^{L-2} + (x-1)^{L-2}\Bigr] x^{-L/2} \;,
$$
which is Eq.\ \eqref{expect_2}.

We now compare Eq.\ \eqref{expect_2} to the internal energy of the system.  The
energy is given in general by $E = -d \ln Z/d \beta = (1/Z)dZ/d\beta$.
Using (\ref{Z_2}) and $dx/d\beta = 2Jx$ (since $x = e^{2\beta J}$), we obtain
from the product rule
\begin{widetext}
\begin{eqnarray}
\label{lengthy}
-E/J &=& \{-(L/2) x^{-L/2-1} 2 x [(x+1)^L + (x-1)^L] + 2x x^{-L/2}[L(x+1)^{L-1} + L(x-1)^{L-1}]\}/Z \nonumber\\ 
	  &=& \{-Lx^{-L/2} [(x+1)^L + (x-1)^L] + 2Lx x^{-L/2}[(x+1)^{L-1} + (x-1)^{L-1}]\}/Z \nonumber\\
	  &=& Lx^{-L/2}\{2x[(x+1)^{L-1} + (x-1)^{L-1}]-[(x+1)^L + (x-1)^L]\}/Z \nonumber\\
	  &=& Lx^{-L/2}\{[2x-(x+1)](x+1)^{L-1} +[2x-(x-1)](x-1)^{L-1}\}/Z\\
	  &=& Lx^{-L/2}\{(x-1)(x+1)^{L-1}+(x+1)(x-1)^{L-1}\}/Z \nonumber\\
	  &=& Lx^{-L/2}\{(x-1)(x+1)(x+1)^{L-2}+(x+1)(x-1)(x-1)^{L-2}\}/Z \nonumber\\
	  &=& L(x^2-1)x^{-L/2}\{(x+1)^{L-2}+(x-1)^{L-2}\}/Z\;,\nonumber
\end{eqnarray}
\end{widetext}
\vspace*{-3mm}
which simplifies in the thermodynamic limit $L \rightarrow \infty$ to Eq.\ \eqref{energy}.

Comparing (\ref{lengthy}) with (\ref{expect_2}), one readily sees that
\begin{equation}
-\varepsilon = 1 - R
\end{equation}
\vspace*{-2mm}
or
\begin{equation}
R = 1 + \varepsilon
\end{equation}
is true for {\em all} lattice sizes $L$, i.e., the relation between the acceptance rate $R$ for
the Metropolis update and the reduced energy density $\varepsilon$ of the 1D Ising model
does {\em not\/} depend on the length $L$ of the one-dimensional chain with periodic boundary 
conditions.

\subsection{Acceptance rate of the heat-bath algorithm}
\label{subsec-heat_bath_1D_ising}
The expected value of the acceptance probability 
$R$ 
of the heat-bath
update can be calculated similarly to Eqs.\ \eqref{expect_1} and \eqref{expect_2}.
Here we directly compute the acceptance probability: 
acceptance probability of an elementary update of the spin $S_i$ is again defined
by $Q \equiv Q_i + Q_{i-1}$, and the analog of Eq.\ \eqref{expect_1} is
\begin{multline}
R = \sum_{l=0}^{L/2} \Bigl(%
  \frac{1}{1 + x^2} \frac{2l}{L} \frac{2l-1}{L-1}  \\
+ \frac{x^2}{1 + x^2} \frac{L-2l}{L} \frac{L-2l-1}{L-1} \\
+ \frac{L-2l}{L} \frac{2l}{L-1}
\Bigr)
\frac{\comb{L}{2l} x^{2l} 2 x^{-L/2} }{Z} \;,
\label{expect_1_HB}
\end{multline}
where the terms in brackets correspond to $Q=2$, $Q=0$ and $Q=1$, respectively.
Differentiating the binomial formula, we obtain 
\begin{equation}
R = \frac{x}{1 + x^2} \frac{1 - \kappa^L}{1 + \kappa^L}\;,
\label{acpt_HB-in-x}
\end{equation}
%
where $\kappa = (x-1)/(x+1) = e^{-1/\xi} < 1$ with $\xi$ denoting the
correlation length. In the thermodynamic limit $L \rightarrow
\infty$, the second factor $\frac{1 - \kappa^L}{1 + \kappa^L} =
\frac{1 - e^{-L/\xi}}{1 + e^{-L/\xi}}$
in \eqref{acpt_HB-in-x} approaches unity exponentially fast, and comparing 
to \eqref{energy}, we obtain Eq.\ \eqref{acpt_HB}.


\end{document}